\documentclass[11pt]{article}

\usepackage{paper}
\usepackage{subcaption}

\title{\textbf{Perspicacious: An Educational Intelligence Platform\\
  for Instructional Designers}}

\author{Yumou Wei\textsuperscript{*}}
\author{Michael C Sutton\textsuperscript{*}}
\author{Paulo F. Carvalho}
\affil{Human-Computer Interaction Institute, Carnegie Mellon University,
  Pittsburgh, PA, USA\\
  \fontshape{n}\selectfont\{yumouw, msutton2, pcarvalh\}@andrew.cmu.edu}

\runningtitle{Perspicacious: An Educational Intelligence Platform}

\begin{document}

\maketitle
{\renewcommand{\thefootnote}{\fnsymbol{footnote}}%
 \footnotetext[1]{Equal contribution.}}

\begin{abstract}
Designing effective instruction requires instructional designers to create assessment questions aligned with clearly defined knowledge components (KCs). These KCs represent specific knowledge or skills that students must acquire. This process is often time-consuming and relies heavily on expert labor. Although AI for education research has produced a growing body of automated question-generation and KC-discovery methods, few have moved beyond research prototypes into user-friendly tools that instructional designers can readily use. To bridge the research-to-practice gap, we developed and evaluated an educational intelligence platform that unifies question generation and KC discovery behind a single interface to support instructional content development. Named ``Perspicacious,'' our platform aims to help instructional designers ``see through'' complex data and algorithms to derive instructional insights. A cost model calibrated on 17 production runs revealed a cost-time trade-off: extracting KCs from a 576-question bank cost US \$1.98--4.11 and took 36--158 minutes to complete, depending on the budget. A formative, partially Wizard-of-Oz study with eight instructional designers found strong perceived usability (mean SUS 78.8) and surfaced concrete requirements that we treat as prerequisites for deployment.

\par\vspace{0.75em}\noindent\textbf{Code:}\enspace
\url{https://github.com/the-oak-lab/KClusterInterface}
\end{abstract}

\section{Introduction}
\label{sec: intro}

Recent advances in AI have reshaped and reinforced the role of education in society. While today's AI can automate many routine tasks, research-backed instructional practices are still needed to develop minds that think critically and creatively. Among the key stakeholders involved in education, \textbf{instructional designers} play a crucial role in delivering effective instruction: they analyze learning needs, develop instructional content, and evaluate learning outcomes~\citep{branch2009}---often behind the scenes. In particular, developing instructional content is a key process that requires two related tasks: (1) generating high-quality assessment questions based on high-level learning objectives and (2) categorizing these questions into knowledge components (KCs)~\citep{Koedinger2012} that represent discrete units of knowledge. Until recently, these tasks required deep content and instructional expertise, making them difficult to scale.

Scaling these tasks to support personalized instruction is often desired because the theory of mastery learning~\citep{bloom_learning_1968} predicts that most people can develop expertise in any domain if the instruction sets personal milestones. However, developing instructional content has traditionally been a laborious process highly dependent on manual work~\citep{stamper_human-machine_2011}. In a typical workflow, instructional designers first draft assessment questions and then engage subject experts in a think-aloud study~\citep{Ericsson1993} to identify the fine-grained KCs required to answer each question. This process is considered the gold standard for developing KC-aligned instructional content, but it does not scale well to large courses---not to mention the many fluid subjects on frontier AI techniques (e.g., agentic engineering) where new skills emerge rapidly. These courses often require frequent updates and continuous development to keep up with the evolving knowledge and the growing demand for personalization. This places a heavy burden on instructional designers who rely largely on manual processes.

Fortunately, decades of research have yielded a wealth of methods and tools that can automate question generation~\citep{Bulathwela2023,Scaria2024,wei2026} and KC discovery~\citep{Shabana2023,Moore2024,wei2025}, with the vision of significantly reducing the workload for instructional designers. Although these methods have been rigorously validated on benchmark datasets or in lab experiments, \textbf{most remain research prototypes that are not yet ready to integrate into instructional design workflows}. If an instructional designer needed to create and categorize thousands of questions by tomorrow, their best bet would still be ``asking ChatGPT''; very few would read the latest research papers and try to apply the cutting-edge methods, only to stop at configuring the runtime environment.

This mile-wide gap between research and practice exists because researchers and instructional designers have different mindsets and priorities. Researchers often focus on building prototypes that improve performance on specific metrics, with no particular consideration for real-world deployment. In contrast, instructional designers prefer intuitive and out-of-the-box tools that seamlessly integrate into their existing workflows without requiring extensive technical expertise. As a result, improved algorithmic performance does not always translate into practical benefits for instructional designers, especially if it raises the technical barrier to use. \textbf{For research output to benefit instructional designers, a user-centric integration is needed to streamline the last-mile adoption of AI tools in instructional design workflows}.

In this paper, we introduce \texttt{Perspicacious}, an AI-powered educational intelligence platform that aims to bridge the gap. The name \texttt{Perspicacious} reflects our vision of providing instructional designers with a clear and insightful view of the instructional design process by helping them ``see through'' the intricate student data and modeling algorithms involved---much like how business-intelligence tools help analysts interpret transaction data. The platform aims to integrate instruction-related information, tools, and resources into a unified interface to streamline existing workflows---including question generation and KC discovery. We describe the design considerations, system architecture, and cost analysis of \texttt{Perspicacious}, and report the results of a formative user study with eight instructional designers. Our findings highlight the importance of usability, seamless integration into existing workflows, and clear explanations of AI-generated outputs to foster trust and adoption among instructional designers.

\section{Related Work}
\label{sec: related}

Generating high-quality questions and identifying KCs from questions are two central tasks in instructional design. We briefly review the literature on AI-assisted question generation and KC discovery, highlighting the progress made in both areas, as well as the existing research-to-practice gap that \texttt{Perspicacious} aims to bridge.

AI-assisted question generation has been an active research area for decades~\citep{kurdi2020}. Early approaches include rule-based methods~\citep{heilman2010,Adamson2013,Wang2008} and sequence-to-sequence models using recurrent neural networks~\citep{du2017,duan2017}. More recently, LLMs have become the de facto standard for question generation. A large body of work~\citep{blobstein2023angel,Hwang2024,Scaria2024} focuses on developing effective prompting strategies to generate questions that align with the revised Bloom's taxonomy of cognitive complexity~\citep{anderson2001taxonomy}. Some recent work has also explored the use of smaller language models for question generation, by fine-tuning~\citep{Bulathwela2023,Li2025} or extensive generation~\citep{wei2026}.

A knowledge component (KC) was first conceptualized as ``an acquired unit of cognitive function or structure that can be inferred from data''~\citep{Koedinger2012}, but the idea of categorizing questions into discrete units of knowledge can be at least traced back to Knowledge Tracing~\citep{corbett_knowledge_1994}. KCs are important building blocks of adaptive learning systems that support mastery learning~\citep{bloom_learning_1968} and are fundamental to evidence-based instructional design that derives educational insights from student data.

AI and machine learning have been the backbone of many automated KC discovery methods. Many automated methods require student data to provide supervision signals. Early approaches include the Q-matrix method~\citep{barnes_q-matrix_2005} and its sequels based on matrix factorization~\citep{desmarais_item_2012,desmarais_matrix_2013}, which search for an optimal question-KC mapping that best predicts student performance. Learning Factors Analysis~\citep{cen_learning_2006} instead searches with A$\ast$~\citep{russell_artificial_2020} over merge and split operations on an existing KC model. Deep-learning-based variants learn the question-KC mapping end-to-end, either from the hidden states of a correctness predictor~\citep{shi_kc-finder_2023} or by extending the Q-matrix with a recurrent architecture~\citep{pardos_dafm_2018}.
LLMs have been applied to KC discovery without requiring student data. Examples include methods that simulate instructional experts~\citep{Moore2024}, extract multimodal signals~\citep{moon2025}, and use next-token probabilities to detect patterns of question co-occurrence~\citep{wei2025}.

Instructional-design platforms exist at either end of the workflow. DataShop~\citep{koedinger2010datashop} gives researchers tools to fit and compare KC models, but presupposes the availability of student data; authoring tools such as CTAT~\citep{aleven2009} let non-experts build and deliver tutors, but take the questions and the KC model as given. \texttt{Perspicacious} sits between them, producing both from a list of learning objectives or an existing question bank.

Despite the impressive research progress so far, a significant research-to-practice gap remains. Very few (if any) of these automated methods have evolved from a research prototype into a user-friendly tool that instructional designers can readily use. Worse, most require more technical expertise than a typical instructional designer has.
For example, even the seemingly simple approach to extract KCs using an LLM involves elaborate prompt engineering or fine-tuning that may overwhelm non-experts, not to mention that LLMs tend to produce an excessive number of KCs with near-duplicates~\citep{Moore2024}. \texttt{Perspicacious} allows instructional designers to upload questions in various formats and receive a KC report \emph{without ever knowing its internals}---as laid out in its user-centric design considerations.

\section{Design Considerations}
\label{sec: design}

Learning to use a new technology can cause stress and anxiety~\citep{Brod1984}, even if it increases productivity. Instructional designers who are already managing a heavy workload may feel overwhelmed by a new platform, and the fear of making mistakes or not understanding it defeats the purpose of adopting it. We therefore grounded \texttt{Perspicacious} in three user-centric design considerations.

\subsubsection*{Low Technical Barriers}
Although we expect our users to become more familiar with AI as more AI literacy programs unfold, using \texttt{Perspicacious} should not require any prior knowledge of AI or machine learning. The platform is designed to require minimal technical expertise. Users can upload a simple plain-text file of learning objectives to generate questions or upload existing questions in various formats (e.g., CSV, XLSX, or JSON) to receive a report mapping each question to a KC. The system handles all intermediate processing while keeping users informed about the progress.

\subsubsection*{Asynchronous Workflows}
Instructional designers often have busy schedules and may be reluctant to devote a lot of time to \texttt{Perspicacious}, so the platform supports asynchronous workflows. Users upload files at their convenience and receive notifications when the job is done, allowing them to manage other duties and easing the fear that a new tool will disrupt their existing workflows.

\subsubsection*{Flexible Scalability}
Users have varying expectations of the service quality. When analyzing a question bank, some users may prefer a minimal turnaround time regardless of the cost, while others may be more sensitive to budget constraints and willing to accept a longer processing time for lower costs. \texttt{Perspicacious} must adjust to different scalability requirements based on the user's expected service quality.

\section{System Architecture}
\label{sec: architecture}

The design considerations led us to build \texttt{Perspicacious} as a web application supported by a modular architecture. The web interface facilitates user interactions with two innovative, LLM-based components that would have been difficult for instructional designers to use directly: a \textbf{KC extraction engine}, which is deployed and characterized in
Section~\ref{sec: cost}, and a \textbf{question generation engine}, which is
implemented but not yet integrated into the production pipeline. We describe the algorithms behind each engine and the web interface. More implementation details are described in Appendix~\ref{app: vertex}.

\subsection{KC Extraction Engine}

The KC extraction engine is a more scalable implementation of KCluster~\citep{wei2025}, a novel KC extraction algorithm based on an LLM. Given a question bank, KCluster identifies clusters of similar questions that potentially belong to the same KC. It quantifies question similarity based on a novel measure of \emph{question co-occurrence}. For every pair of questions, $q_{s}$ and $q_{t}$, KCluster uses \emph{Phi-2}~\citep{hughes_phi-2_2023} as a native \emph{probability machine} to evaluate the change in log-probability of $q_{s}$ given $q_{t}$:
\begin{equation}\label{eqn: pmi}
\mathrm{\Delta} (q_{s}, q_{t}) := \log \Pr(q_{s} \mid q_{t}) - \log \Pr(q_{s})
\end{equation}
and defines a symmetric measure of question similarity:
\begin{equation}
\text{Congruity}(q_{s}, q_{t}) := \frac{1}{2} \left[ \mathrm{\Delta} (q_{s}, q_{t}) + \mathrm{\Delta} (q_{t}, q_{s}) \right]
\end{equation}

The idea is that if the occurrence of $q_{t}$ significantly increases the probability of $q_{s}$ occurring (as measured by the conditional, $\log \Pr(q_{s} \mid q_{t})$) over the baseline occurring rate $\log \Pr(q_{s})$, then $q_{s}$ and $q_{t}$ are congruent---they may come from the same textbook chapter, concern the same topic, or contain similar texts, all suggesting that they belong to the same KC. On the other hand, if $q_{t}$ barely changes or even decreases the likelihood of $q_{s}$ occurring, then $q_{s}$ and $q_{t}$ are incongruent and likely to belong to different KCs. KCluster then applies a clustering algorithm to systematically determine whether two questions belong to the same KC based on the congruity between all pairs of questions.

In the original formulation of KCluster, it requires $N$ evaluations to extract the concept labels for all $N$ questions and $N^{2} + N$  evaluations to calculate the congruity for every question pair---a total of ${\mathrm{\Omega}}(N^2)$ evaluations that do not scale well if $N$ is large. To support asynchronous workflows and flexible scalability for \texttt{Perspicacious}, we re-implemented KCluster on Google's Vertex AI platform to leverage its powerful distributed computing capabilities and asynchronous processing features (see Appendix~\ref{app: vertex}). In addition, Vertex AI allows us to estimate the computing resources to allocate based on a user's desired completion time to accommodate workloads flexibly (more details in Section~\ref{sec: cost}).

To illustrate the engine's output, the 64 questions generated for our user study (Section~\ref{sec: user_study}) produced eight KCs, and 63 of the 64 questions were grouped with the other questions written for the same objective (see Table~\ref{tab: kc-example} in the appendix). The exception, ``\emph{What is the carbon cycle?}'', was placed with questions about carrying capacity. Because the questions were generated from those objectives, this is only a sanity check; on real question banks, KCluster's KC models predict student performance better than the best expert-designed models available~\citep{wei2025}.

\subsection{Question Generation Engine}
The question generation engine replicates a recent ``Generate-Then-Validate'' pipeline~\citep{wei2026} that uses an LLM to create high-quality questions from learning objectives. The pipeline proceeds in two main stages: expansive generation and selective validation.

In the expansive generation stage, Phi-2 is prompted to generate a large number of candidate questions for each learning objective, regardless of question quality. These questions are then validated and filtered through a three-step validation process. The first step is syntactic filtering, which removes any question that violates basic syntactic rules (e.g., containing incomplete options).
Next, Phi-2 is used to evaluate the probability of the answer key given each question, and questions with low answer-key probabilities ($\leq 0.9$) are discarded. Finally, a similar notion of change in log-probabilities is extended to question-objective pairs to measure the relevance of each question ($Q$) to each learning objective ($L$):
\begin{equation}
    \text{Relevance}(Q, L) := \log \Pr(Q \mid L) - \log \Pr (Q)
\end{equation}
The idea is again that if a learning objective elicits a large increase in log-probability of a question over its baseline occurring rate, the learning objective is likely to be relevant to the question. The pipeline removes any question whose original learning objective does not yield the highest relevance score among all competing learning objectives.

\subsection{Web Interface}

Users need to know none of these technical details thanks to the user-friendly web interface. The web interface is built with Django and deployed on Google Cloud Run in a Docker container. After signing in, users see a dashboard (Figure~\ref{fig: dashboard} in the appendix) summarizing their past and present tasks, from which they can track progress, download the generated questions or KC report, or run further KC analyses.

The ``New Analysis'' button opens the upload page (Figure~\ref{fig: upload} in the appendix). For question generation, users can upload a plain-text file containing the target learning objectives. For KC extraction, we accept a wider range of formats that question banks may already come in, whereas the original KCluster algorithm requires a specific JSONL input format.
To minimize user effort, \texttt{Perspicacious} accepts Microsoft Excel (.xlsx), CSV (.csv), and JSON (.json) formats, and automatically converts them into the correct JSONL structure. To reduce formatting errors, the interface provides (1) \textbf{clear field descriptions} explaining what information is required, (2) \textbf{example files} showing correctly formatted data, and (3) \textbf{downloadable templates} for each supported format. These aim to ensure that even first-time users can prepare correct input without extensive trial-and-error.
Both engines are hosted inside the deployment's own Google Cloud project, so uploaded questions never reach a third-party LLM service; retention and access policies for those uploads remain the hosting institution's to set.

\section{Cost and Scalability Analysis}
\label{sec: cost}

The cost of running \texttt{Perspicacious} is determined by the computing resources allocated to a task: more resources shorten the turnaround time but raise the cost. To meet the varying expectations as described in our design considerations (Section~\ref{sec: design}), \texttt{Perspicacious} can flexibly allocate resources based on the user's desired completion time.

We analyze the cost and scalability of the KC extraction engine. The engine uses Phi-2, a lightweight 2.7B-parameter LLM, which we host on the most affordable \texttt{g2-standard-4} instances on Vertex AI. Each instance includes an NVIDIA L4 GPU with 24GB of memory. At the time of writing (August 2026), the cost of running one such instance is US \$0.8129 per hour. If the \emph{actual} completion time is $A$ minutes, the actual cost $C$ is given by:
\begin{equation}\label{eqn: cost}
    C = 0.8129 \cdot \frac{A}{60} \cdot I
\end{equation}
The number of instances to allocate ($I$) can be determined by the user's desired completion time. Let $N$ be the number of questions uploaded by the user. As mentioned previously, the total number of predictions to make is $N^{2} + 2N$, which includes $N$ predictions for concept labels and $N^{2} + N$ predictions for $\Delta(q_{s}, q_{t})$ values. Vertex AI processes these predictions in batches of size $B$, which we typically set to 16 to fully utilize the GPU memory available. Each batch takes $S$ seconds to process, which we estimate empirically; overestimating $S$ over-provisions the job and inflates cost, while underestimating it starves the job and lengthens processing. Vertex AI also documents that each instance requires some time $T_{0}$ (15 minutes or so) to set up before any request is served. If the desired completion time is $T$ minutes, the number of instances to allocate can be estimated by:
\begin{equation}\label{eqn: alloc}
    I = \max \left(1, \left\lceil\frac{(N^{2} + 2N) \cdot S}{B \cdot (T - T_{0}) \cdot 60}\right\rceil\right),\;\; T > T_{0}
\end{equation}
The actual completion time $A$ varies with system load and resource availability: across runs at matched $N$ and $I$ with different question banks, completion times agreed to within 4.4--14.6\%, and two question banks at $I = 1$ agreed to 8.1\%.

In practice, a user uploads $N$ questions and indirectly sets the number of instances $I$ via their desired completion time $T$. A good measure of the system workload $W$ is then \emph{batches per instance}, $W = (N^{2} + 2N) / (B \cdot I)$, which accounts for the different budget constraints conveyed by $I$. To understand how the actual completion time $A$ relates to $W$, we fit a linear regression model $A \sim T_{0} + \frac{S}{60}\cdot W$ to the job completion statistics of 17 jobs. These jobs were completed for a previous project that extracted KCs from 13 datasets curated by~\citet{cao2026what}, spanning science courses at four schools.

\begin{figure}[t]
    \centering
    \begin{subfigure}[t]{0.48\textwidth}
        \centering
        \includegraphics[width=\linewidth]{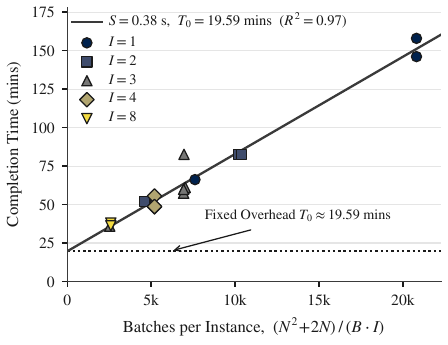}
        \caption{Actual completion time against batches per instance, over 17
            production KC-extraction runs. The fit gives $S \approx 0.38$ seconds per batch
            and a fixed overhead of $T_{0} \approx 19.59$ minutes.}
        \label{fig: cost-calibration}
    \end{subfigure}\hfill
    \begin{subfigure}[t]{0.48\textwidth}
        \centering
        \includegraphics[width=\linewidth]{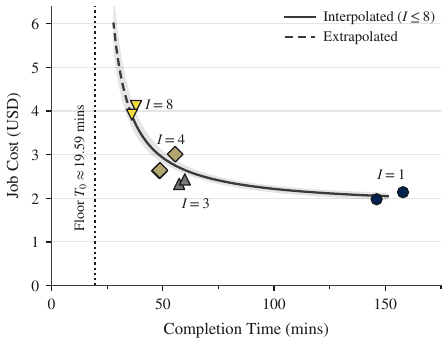}
        \caption{Cost against actual completion time for a 576-question bank. Markers are the nine measured runs, each at a given $I$; the curve is drawn solid over the measured $I$ range and dashed beyond it.}
        \label{fig: cost-tradeoff}
    \end{subfigure}
    \caption{Cost and scalability analysis of the KC extraction engine.}
    \label{fig: cost}
\end{figure}

Figure~\ref{fig: cost-calibration} shows that the linear model fits well to the data ($R^{2} = 0.97$) across different $I$. It estimates the processing speed $S \approx 0.38 \pm 0.02$ seconds per batch and the fixed start-up overhead $T_{0} \approx 19.59 \pm 2.67$ minutes, some 30\% above the documented figure; we use this fitted value in Equation~\ref{eqn: alloc}. Substituting $A$ in Equation~\ref{eqn: cost} with the estimated linear model gives an informative decomposition of the actual cost $C$:
\begin{equation}
    C = \underbrace{\frac{0.8129\cdot S}{60\cdot 60}}_{\text{\$ per batch}} \cdot \underbrace{\frac{N^{2} + 2N}{B}}_{\text{num of batches}} + \underbrace{\frac{0.8129\cdot T_{0}}{60}}_{\text{start-up overhead}} \cdot \,\, I
\end{equation}

At $N = 576$, the median and mode size of the 13 datasets analyzed, this simplifies to $C = 1.78 + 0.265I$. In other words, extracting KCs from 576 questions costs an irreducible \$1.78 plus an overhead proportional to $I$. Under the same linear model, the actual cost $C$ is a rational function of $A$ with a vertical asymptote at $T_{0}$: $C \propto A / (A - T_{0})$. Figure~\ref{fig: cost-tradeoff} shows such a cost curve at $N = 576$ across nine jobs run at various $I$. We observe a diminishing return for adding more instances as $A$ approaches the irreducible floor at $T_{0}$. Extracting KCs from a 576-question bank cost between \$1.98 and \$4.11 and took 36 to 158 minutes, depending on the budget. After \texttt{Perspicacious} is deployed, we can collect more usage data to better estimate the resources needed for a given job and provide greater cost transparency for users.

\section{User Study}
\label{sec: user_study}

To evaluate its usability and pedagogical utility, we conducted a user testing study of \texttt{Perspicacious}'s ability to enable instructional designers to integrate two existing AI-based methods: KCluster~\citep{wei2025} for KC discovery and Generate-Then-Validate~\citep{wei2026} for question generation.
This study goes beyond algorithmic evaluation to examine real-world user interaction, adoption barriers, and the interpretability of outputs in an instructional context. The study was conducted between January 2026 and June 2026.

\subsection{Participants}
We recruited eight instructional designers from within our professional network. All participants regularly work with assessment questions and have varying levels of experience with tools that extract KCs and generate questions. Although participants had professional experience in instructional design, they were not domain experts in the specific subject area used in the study materials; accordingly, their evaluations focused on usability, alignment, and pedagogical plausibility, rather than on fine-grained content accuracy. We refer to our participants as P1--P8 in order of session date.

\subsection{Procedure}
Participants were invited to use the platform in a think-aloud session lasting about 40 minutes. They completed two tasks:
\begin{enumerate}
    \item Upload a plain-text file of eight learning objectives to generate questions. The learning objectives were chosen from the same Environmental Science curriculum used by~\citet{wei2026}.
    \item Download a CSV file of the 64 generated questions and re-upload it to generate KCs for each question.
\end{enumerate}
After each task, participants reviewed a subset of the generated questions and KC assignments and rated them in terms of clarity and perceived usefulness. Finally, participants completed a post-study survey consisting of usability questions and additional open-ended questions asking how they might integrate the platform into their workflows, as well as any remaining challenges or points of confusion. All experimental procedures were approved by our institution's IRB.

\subsection{Evaluation Scope}
\label{sec: scope}
Our user study evaluated the interface and the end-to-end workflow rather than engine performance, and was conducted as a partial Wizard-of-Oz evaluation~\citep{kelley1984,dahlback1993}. Account creation, file upload and validation, format conversion, the task dashboard, and email notification were live. The questions and KC assignments returned during a session, however, were prepared in advance with the same Generate-Then-Validate~\citep{wei2026} and KCluster~\citep{wei2025} pipelines rather than computed in real time, because a KC extraction job takes about 20 minutes to \emph{start} (Section~\ref{sec: cost})---too long for two think-aloud tasks, output review, and a survey to be done in a 40-minute session. The question generation engine was not yet integrated. Participants were explicitly told that parts of the system were simulated and that outputs might not reflect the fully integrated version. The results below therefore characterize the workflow and practitioners' judgments of output quality, not end-to-end latency or reliability when the platform goes live.

\subsection{Measures}
Quantitative measures included task completion rate and usability scores, assessed using the 10-item System Usability Scale (SUS)~\citep{brooke1996sus}. Qualitative data consisted of thematic analyses of interviews focusing on ease of use, clarity of outputs, and suggestions for improvement.

\subsection{Results}

All eight participants completed both tasks. For P4, the dashboard did not update before time ran out and the facilitator supplied the prepared output; P8 met an unsupported file format and recovered by exporting to CSV. In the post-study survey, seven of the eight participants anticipated using the platform in their own work, while one did not.

\subsubsection*{Usability}
Participants rated each SUS item on a five-point Likert scale, and we scored the items according to the standard procedure (see Appendix~\ref{app: sus-administration}).
For our study, participants found the web application easy to use and learn (raw scores 26--37, corresponding to standardized SUS scores of 65--92.5; M = 78.8, SD = 10.3). The average score falls in the ``B+'' range on the curved grading scale of SUS benchmarks, indicating good perceived ease of use and learnability~\citep{lewis2018sus}. Figure~\ref{fig: sus} in the appendix shows the overall and item-level score distributions.

\subsubsection*{Qualitative Feedback}
To understand the perceived utility of the platform, we analyzed the think-aloud transcripts and post-session surveys of the eight instructional designers. Our findings center on participants' existing workflows, the friction in current AI-assisted practices, the platform's practical utility, and the conditions they require for adoption.

\begin{enumerate}
    \item \textbf{Business as Usual}
        \\ Participants described assessment design as a highly manual process under persistent time constraints. They use general LLMs such as ChatGPT and Gemini for brainstorming, but conversationally rather than integrated into established workflows. P3 reported they ``\emph{use ChatGPT every week for brainstorming questions},'' while P5, a learning engineer, wished their team had ``\emph{a system better than generic ChatGPT}.'' These ad hoc uses have limits: generic LLMs produced ``\emph{trivial distractors}'' at scale (P3), requiring item-by-item revision, and some participants wanted more robust ways to derive KCs from questions of varying quality. Practitioners struggle not with access to LLMs but with the lack of tools that integrate generation, validation, and refinement into a coherent workflow.
    \item \textbf{Platform Utility and Usability}
    \\ Participants responded positively to the platform's ability to provide access to validated assessment-generation methods through a simple visual interface. Rather than requiring Python scripts or command-line tools, practitioners could directly invoke the Generate-Then-Validate pipeline and obtain assessment items they perceived as more rigorous than typical conversational LLM output. P8, who creates assessment items daily, remarked that ``\emph{this would be very helpful},'' while P2 described the interface as ``\emph{clear and nice from a usability standpoint}.'' Participants also valued the quality of generated answer choices and distractors, noting that they appeared less superficial than those produced by ChatGPT and could support meaningful instructional discussion. The low technical barrier was similarly emphasized, with P3 reporting that they could have learned to use the system independently within minutes because the templates were easy to understand.
    \item \textbf{Trust as a Precondition for Adoption}
    \\ Two participants made adoption conditional on trusting the output. P1 would use the platform ``\emph{if I can trust the quality},'' naming distrust of ``\emph{the generated results}'' as the only reason they would not; P2 cited distrust of ``\emph{AI hallucination}'' and asked to ground generation in uploaded source material. Neither engine currently exposes why a question or KC was produced, suggesting that provenance and verification are as important for adoption as usability.
\end{enumerate}

\section{Discussion}
\label{sec: discussion}

Participants' strong response to low-barrier access to the Generate-Then-Validate and KCluster pipelines suggests that the research-to-practice gap is not just a question of algorithmic performance, but also of whether effective methods can be embedded in workflows that practitioners can adopt.

\subsection{Improving Cognitive Fidelity of KCs}
Five participants noted that the generated KCs all used the same action verb ``identify'' (see Table~\ref{tab: kc-example} in the appendix). Although the underlying concepts were generally judged to be appropriate, this repetition reduced the usefulness of the KCs for distinguishing between lower- and higher-level learning goals. Using a wider range of action verbs could make the generated KCs more informative for instructional planning and better support alignment with cognitive frameworks such as the revised Bloom's taxonomy~\citep{anderson2001taxonomy}.

\subsection{Requirements for Deployment}

Deployment is more within reach than what our prototype may have suggested. The KC extraction engine already runs in production on Vertex AI---the 17 jobs analyzed in Section~\ref{sec: cost} processed 13 real question banks from four schools---and account creation, upload, validation, the dashboard, and notification all ran live in the study. Integrating the question generation engine is the one remaining engineering step, and our participants set four requirements for that release.

The first is steerability. Participants valued automated generation while requesting control over difficulty, reading level, question type, and item refinement: \textbf{practitioners appear to want more steerable automation rather than autonomous automation}. The goal is not to remove the instructional designer, but to accelerate routine work while preserving their agency. The second is expressiveness: several wanted explanatory, option-specific feedback deliverable to students, and others wanted formats beyond multiple choice (e.g., open-ended, scenario-based, short-answer, and essay questions) to cover a wider range of learning objectives. The third is provenance. Because adoption was conditional on trust, the release must show why each question or KC was produced, so that instructional designers can verify rather than assume.

The fourth comes from our cost model. Because every job pays a fixed start-up overhead of about 20 minutes (Section~\ref{sec: cost}), small jobs are dominated by provisioning: the model puts the study's 64-question bank at 21 minutes on a single instance, 92\% of it spent waiting for that instance to start, against 13\% for a full 576-question course bank. This is precisely the scale at which a new user is likely to try the platform, so keeping a warm instance available, or batching small jobs together, would make first contact with \texttt{Perspicacious} feel immediate rather than slow.

\section{Conclusion and Future Work}
\label{sec: conclusion}

We introduced \texttt{Perspicacious}, an educational intelligence platform that supports both automated question generation from learning objectives and KC extraction from existing assessment items, extending KCluster and Generate-Then-Validate into an end-to-end workflow with low technical barriers for instructional designers. A formative user study demonstrated strong usability and positive perceptions of the platform's pedagogical utility.
Future work will focus on refining the system based on user feedback, such as increasing control over question generation parameters and enhancing KC expressiveness, and on follow-up evaluations with subject experts and classroom instructors to assess content accuracy and instructional impact in authentic educational settings.

\section*{Acknowledgments}
This research was supported by the US National Science Foundation under Grant No. 2301130 and a Google Academic Research Award to Paulo F. Carvalho.

\printbibliography

\appendix

\section{System Implementation}
\label{app: vertex}

\begin{figure}[t]
    \centering
    \includegraphics[width=0.9\textwidth]{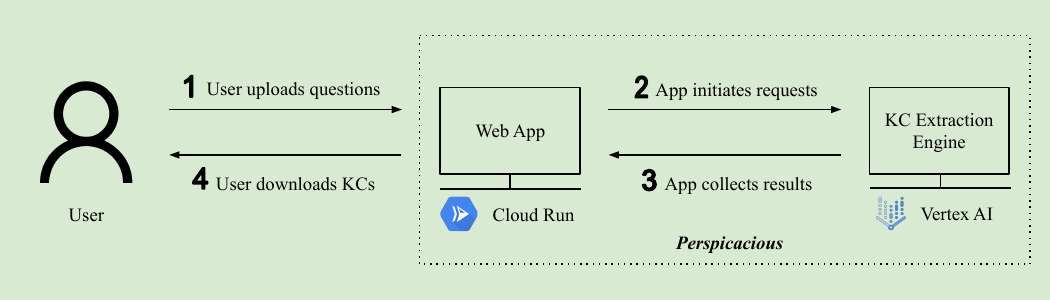}
    \caption{An example workflow initiated by a user to extract KCs from questions}
    \label{fig: user_flow}
\end{figure}

\begin{figure}[t]
    \centering
    \includegraphics[width=0.9\textwidth]{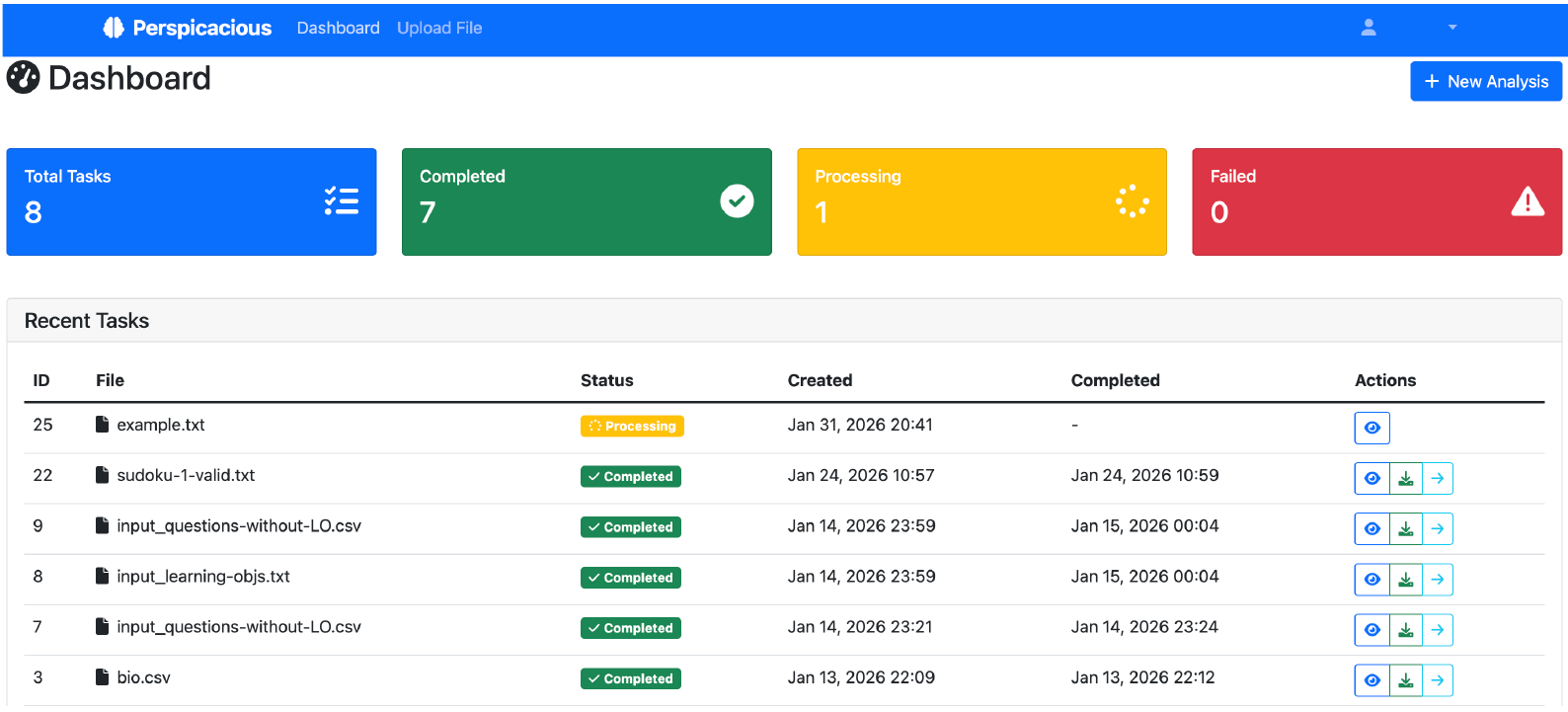}
    \caption{An intuitive dashboard showing an overview of a user's KC extraction tasks}
    \label{fig: dashboard}
\end{figure}

The \texttt{Perspicacious} web application is hosted on Google Cloud Platform. When a user uploads a file, it is uploaded to a designated bucket on Google Cloud Storage, and a Cloud Run job corresponding to a task on the dashboard (Figure~\ref{fig: dashboard}) is then created to begin the appropriate task. The job first verifies that the uploaded file satisfies the requirements and then converts the file into the required input format, which is then processed and sent to the appropriate (KC extraction or question generation) engine hosted on Vertex AI (now named ``Gemini Enterprise Agent Platform''). A relational database is maintained via Cloud SQL to store user and task information. The web application uses multithreading to monitor the status of each task. Once a task is completed, the user is notified via email to download the generated questions or KCs. Figure~\ref{fig: user_flow} illustrates an example workflow when a user initiates a KC extraction task.

\subsubsection*{KC Extraction Engine}
The KC extraction engine runs a more scalable implementation of the KCluster algorithm~\citep{wei2025}. The KCluster algorithm first extracts concept labels from each question and then computes $\Delta(q_{s}, q_{t})$ in Equation~\ref{eqn: pmi} for every pair of questions $q_{s}$ and $q_{t}$. With reference to the original implementation\footnote{\url{https://github.com/weiyumou/KCluster}}, we built a Docker image that packs label extraction and $\mathrm{\Delta}$ calculation as a Custom Prediction Routine
and registered it as a Model on Vertex AI. When a user uploads $N$ questions, the web application prepares a JSONL file of $N^{2} + 2N$ lines, each of which is a prediction request containing the prompt for either label extraction or $\mathrm{\Delta}$ calculation ($N$ requests for label extraction and $N^{2} + N$ requests for $\mathrm{\Delta}$ calculation). The web application then initiates a BatchPredictionJob to serve the requests. This process is fully managed and optimized by Vertex AI, which applies MapReduce to parallelize the computation and allocate sufficient resources commensurate with the user's expected quality of service. When the batch job is completed, the web application retrieves the KC labels and $\mathrm{\Delta}$ values before applying the clustering algorithm to derive the KCs. The user is then notified to download a CSV report that maps each question to a KC.

\subsubsection*{Question Generation Engine}
Although we are still in the process of fully integrating the question generation engine into \texttt{Perspicacious} (see Section~\ref{sec: scope}), we plan to pack a scalable implementation of the Generate-Then-Validate pipeline~\citep{wei2026} in a Docker image and register it as another Model on Vertex AI, similar to the KC extraction engine. When a user uploads a set of learning objectives, the web application will prepare a JSONL file containing prediction requests for the generation stage. After initiating a {BatchPredictionJob} for the generation stage, the web application retrieves the generated questions and prepares another JSONL file for the validation stage. A second {BatchPredictionJob} is then initiated for the validation stage. Once the validation is completed, the user is notified to download a CSV file containing the final set of questions.

\begin{figure}[t]
    \centering
    \includegraphics[width=0.9\textwidth]{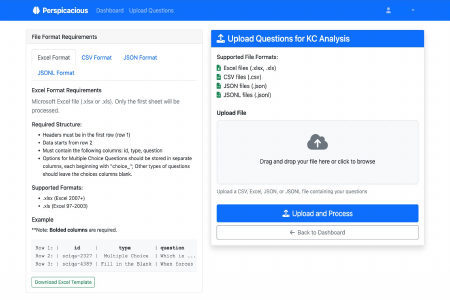}
    \caption{An informative upload page guiding the user to upload questions in a variety of supported formats}
    \label{fig: upload}
\end{figure}

\begin{table}[t]
\centering\small
\setlength{\tabcolsep}{5pt}
\caption{Breakdown of the 64 questions used in the study by the learning objective they were written for and the knowledge component \texttt{Perspicacious} assigned them to. Seven of the eight objectives map onto exactly one KC; only the carbon-cycle objective is split, with a single question (``What is the carbon cycle?'') placed under the carrying-capacity KC.}
\label{tab: kc-example}
\begin{tabular}{@{}p{0.85\textwidth}r@{}}
\toprule
\textbf{Learning objective} \,/\, \emph{discovered knowledge component} & \textbf{Questions} \\
\midrule
\textbf{Define the demographic transition} & \textbf{8} \\
\quad\emph{identify the correct definition of the demographic transition} & 8 \\
\addlinespace[2pt]
\textbf{Describe causes of and problems related to overfishing} & \textbf{8} \\
\quad\emph{identify a likely cause of overfishing} & 8 \\
\addlinespace[2pt]
\textbf{Describe methods for mitigating human impact on forests} & \textbf{8} \\
\quad\emph{identify a method of mitigating human impact on forests} & 8 \\
\addlinespace[2pt]
\textbf{Describe sustainable agricultural and food production practices} & \textbf{8} \\
\quad\emph{identify the environmental and economic benefits of sustainable agricultural and food production systems} & 8 \\
\addlinespace[2pt]
\textbf{Describe the impact of carrying capacity on ecosystems} & \textbf{8} \\
\quad\emph{identify the impact of carrying capacity on ecosystems} & 8 \\
\addlinespace[2pt]
\textbf{Describe the impacts of human activity on wetlands and mangroves} & \textbf{8} \\
\quad\emph{identify the impact of human activity on wetlands and mangroves} & 8 \\
\addlinespace[2pt]
\textbf{Explain the environmental effects of excessive use of fertilizers and detergents on aquatic ecosystems} & \textbf{8} \\
\quad\emph{identify the environmental effects of excessive use of fertilizers and detergents on aquatic ecosystems} & 8 \\
\addlinespace[2pt]
\textbf{Explain the steps and reservoir interactions in the carbon cycle} & \textbf{8} \\
\quad\emph{identify the impact of human activities on the carbon cycle} & 7 \\
\quad\emph{identify the impact of carrying capacity on ecosystems} & 1 \\
\addlinespace[2pt]
\midrule
\textbf{Total} & \textbf{64} \\
\bottomrule
\end{tabular}
\end{table}

\begin{figure}[t]
    \centering
    \includegraphics[width=0.9\textwidth]{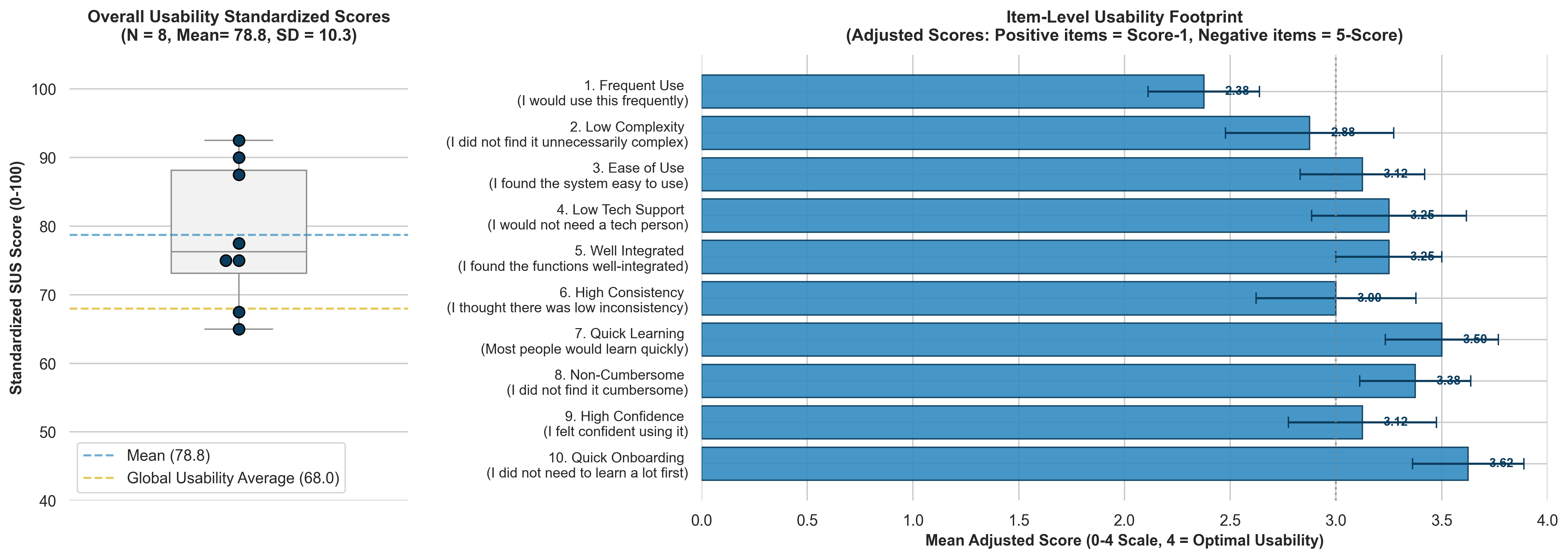}
    \caption{\textbf{System Usability Scale (SUS) evaluation of the platform.}
    The left panel displays the overall distribution of standardized SUS scores (0--100) using a box-and-whisker plot overlaid with individual participant data points. The system achieved a mean standardized SUS score of 78.8 ($\mathrm{SD} = 10.3$), placing it above the industry-average usability baseline of 68.0 (indicated by the dashed gold line). The right panel provides an item-level usability footprint. To allow direct comparison, Likert-scale responses (1--5) are mathematically adjusted to a uniform 0--4 scale (where 4 represents optimal usability) following standard SUS scoring protocols. Error bars represent the standard error of the mean.}
    \label{fig: sus}
\end{figure}

\section{SUS Administration}
\label{app: sus-administration}

Participants completed the System Usability Scale (SUS) following their interaction with Perspicacious. The SUS is a standardized 10-item usability questionnaire. We reproduce the SUS questions administered to participants below for transparency and reproducibility.

The SUS items are sourced from and scored according to the standard SUS procedure described by \citet{brooke1996sus}. Responses to positively worded items were adjusted by subtracting 1, while negative worded item responses were subtracted from 5. We summed the resulting values to obtain the raw scores (0--40) and multiplied them by 2.5 to produce the standardized scores (0--100).

Participants answered the following questions on a Likert Response Scale from 1 to 5, where 1 represents Strongly Disagree and 5 represents Strongly Agree:
\begin{enumerate}
    \item I think that I would like to use this system frequently.
    \item I found the system unnecessarily complex.
    \item I thought the system was easy to use.
    \item I think that I would need the support of a technical person to be able to use this system.
    \item I found the various functions in this system were well-integrated.
    \item I thought there was too much inconsistency in this system.
    \item I would imagine that most people would learn to use this system very quickly.
    \item I found the system very cumbersome to use.
    \item I felt very confident using the system.
    \item I needed to learn a lot of things before I could get going with this system.
\end{enumerate}

\end{document}